\documentclass[conference]{IEEEtran}
\IEEEoverridecommandlockouts
\usepackage{cite}
\usepackage{amsmath,amssymb,amsfonts}
\usepackage{algorithmic}
\usepackage{graphicx}
\usepackage{textcomp}
\usepackage{xcolor}
\usepackage{listings}
\usepackage{url}

\def\BibTeX{{\rm B\kern-.05em{\sc i\kern-.025em b}\kern-.08em
    T\kern-.1667em\lower.7ex\hbox{E}\kern-.125emX}}
\begin{document}

\title{Automated service monitoring in the deployment of ARCHER2}

\author{\IEEEauthorblockN{Mr Kieran Leach}
\IEEEauthorblockA{\textit{EPCC, The University of Edinburgh} \\
Edinburgh, UK \\
k.leach@epcc.ed.ac.uk}
\and
\IEEEauthorblockN{Mr Philip Cass}
\IEEEauthorblockA{\textit{EPCC, The University of Edinburgh} \\
Edinburgh, UK \\
p.cass@epcc.ed.ac.uk}
\and
\IEEEauthorblockN{Mr Steven Robson}
\IEEEauthorblockA{\textit{EPCC, The University of Edinburgh} \\
Edinburgh, UK \\
s.robson@epcc.ed.ac.uk}
\and
\IEEEauthorblockN{Mr Eimantas Kazakevicius}
\IEEEauthorblockA{\textit{EPCC, The University of Edinburgh} \\
Edinburgh, UK \\
e.kazakevicius@epcc.ed.ac.uk}
\and
\IEEEauthorblockN{Mr Martin Lafferty}
\IEEEauthorblockA{\textit{HPE} \\
Edinburgh, UK \\
martin.lafferty@hpe.com}
\and
\IEEEauthorblockN{Dr Andrew Turner}
\IEEEauthorblockA{\textit{EPCC, The University of Edinburgh} \\
Edinburgh, UK \\
a.turner@epcc.ed.ac.uk}
\and
\IEEEauthorblockN{Dr Alan Simpson}
\IEEEauthorblockA{\textit{EPCC, The University of Edinburgh} \\
Edinburgh, UK \\
a.simpson@epcc.ed.ac.uk}
}

\maketitle

\begin{abstract}
The ARCHER2 service, a CPU based HPE Cray EX system with 750,080 cores (5,860 nodes), has been deployed throughout 2020 and 2021, going into full service in December of 2021. A key part of the work during this deployment was the integration of ARCHER2 into our local monitoring systems. As ARCHER2 was one of the very first large-scale EX deployments, this involved close collaboration and development work with the HPE team through a global pandemic situation where collaboration and co-working was significantly more challenging than usual. The deployment included the creation of automated checks and visual representations of system status which needed to be made available to external parties for diagnosis and interpretation. We will describe how these checks have been deployed and how data gathered played a key role in the deployment of ARCHER2, the commissioning of the plant infrastructure, the conduct of HPL runs for submission to the Top500 and contractual monitoring of the availability of the ARCHER2 service during its commissioning and early life.
\end{abstract}


\section{Background}
\subsection{ARCHER and ARCHER2}
In this paper we discuss the deployment and utility of automated monitoring during the recent rollout of the ARCHER2 system and service. The ARCHER2 system is an HPE Cray EX supercomputer with an estimated peak performance of 28 Pflop/s. The machine has 5,860 compute nodes, each with dual AMD EPYC 7742 64-core processors at 2.25GHz, giving 750,080 cores in total. ARCHER2 is the successor system to ARCHER, a 4,920-node Cray XC30 system which was also operated and supported by EPCC. These systems have been managed and financed by the Engineering and Physical Science Research Council (EPSRC) and UK Research and Innovation (UKRI). 

In operating and supporting both these services, EPCC has acted in both the Service Provision (SP) and Computational Science and Engineering (CSE) roles. Under the SP role, EPCC is responsible for system management and administration as well as the operation of the Service Desk. Under the CSE role, EPCC is responsible for deploying application software not included in the programming environment supplied by HPE as well as for assisting users with application software development and management, providing training, administering funding calls for software development projects, and running an outreach programme. These responsibilities are in addition to hosting the ARCHER2 service at EPCC's Advanced Computing Facility (ACF) data centre.
\subsection{Deployment timeline}
Owing to a combination of the impacts of COVID-19 and developmental difficulties with the HPE Cray EX and Slingshot technologies, ARCHER2 experienced an extended and somewhat troubled deployment. The originally planned deployment timeline was:
\begin{itemize}
    \item February 2020: ARCHER to be decommissioned
    \item May 2020: ARCHER2 to be made available to users
\end{itemize}

Given the issues faced with the development and scaling of the HPE Cray EX and Slingshot technologies, it was decided to introduce a phased transition. Instead of a direct transfer to the full 23 cabinet system, a 4 cabinet system was temporarily deployed to a separate computer room, to operate in parallel to ARCHER until such time as it was possible to deploy the full 23 cabinet system. The final deployment timeline was:
\begin{itemize}
    \item July 2020: ARCHER2 4 cabinet system delivered to the ACF
    \item October 2020: ARCHER2 4 cabinet system made available to early access users
    \item November 2020: ARCHER2 4 cabinet system made available to all users
    \item January 2021: ARCHER system decommissioned and removed from the ACF
    \item February 2021: ARCHER2 23 cabinet system delivered to the ACF
    \item November 2021: ARCHER2 23 cabinet system made available to users
\end{itemize}
\subsection{Motivation}
As is discussed further below, when planning the deployment of ARCHER2 in excess of four years experience of using monitoring technologies to improve response time, reduce staff workloads and provide insight when responding to problems. Having had success in integrating monitoring into our approach to service deployment for other HPC services at EPCC, particularly in supporting the diagnosis of problems with services during deployment, monitoring integration is a key part of our standard approach when commissioning a new system. As such, deploying monitoring was integrating into planning for the ARCHER2 deployment from the start.

\begin{figure*}[ht]
  \centering
  \includegraphics[width=1.0\textwidth]{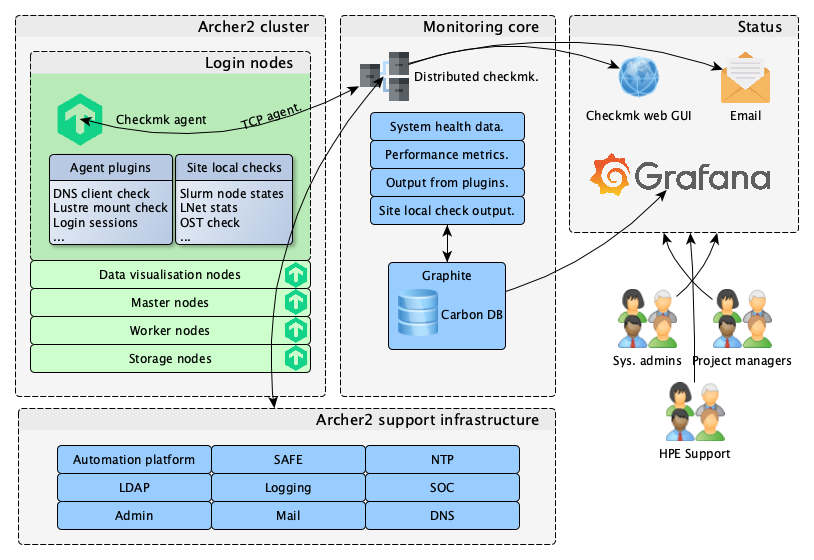}
  \caption{An overview of the deployment of monitoring services for ARCHER2}
  \label{monitoring_setup_diagram}
\end{figure*}
\section{Checkmk and Graphite}
\subsection{General background}

EPCC manages a variety of HPC and research computing services in addition to critical support infrastructure. In the past, EPCC system administrators spent a great deal of time tracking the state of various systems. Problem detection and diagnosis typically required looking in multiple locations and running a variety of monitoring scripts on a regular basis. This was time intensive, complex to operate and maintain, and difficult for team members to manage. This approach also made it problematic to integrate new systems into standard operating procedures as team members are typically under pressure to get services up and running in a short time.

Given all this, it was considered that a "single pane of glass" approach was necessary: using a single screen to monitor the status of all services on site.

The solution selected at EPCC was, and remains, Checkmk\cite{b7}. Originally developed in 2008 as a Nagios extension, Checkmk is now a full Nagios derivative monitoring system. 

Checkmk supports a variety of types of monitoring:
\begin{itemize}
    \item status-based monitoring relating to the health of a system or process;
    \item metric-based monitoring recording data on aspects of a system or process over time; and
    \item log-based monitoring triggering from the detection of events in logs.
\end{itemize}

Among the motivations for selecting Checkmk were:
\begin{itemize}
    \item the range of existing checks including CPU, memory, file system and network interface status; 
    \item the ability to simply deploy new checks;
    \item the ability to simply add new hosts; and
    \item the ability to simply manage and view checks from a single interface.
\end{itemize}

Checkmk was first deployed at EPCC in 2015. Since that time it has become core to the management of our HPC services. It has also allowed us to deploy bespoke monitoring solutions for a variety of HPC technologies.

Our Checkmk dashboard is monitored during working hours by the "on-shift" team member. This has allowed us to gain awareness of and respond to issues quickly; including partial power failures, system, disk and component failures, networking issues and system load issues. Certain alerts are also issued by email. We have made a practice of including stakeholders such as our hardware partners on the list for email alerts for relevant systems, often allowing those stakeholders who work outside our working hours to become aware of and resolve issues before anyone from EPCC enters said working hours. This has included ensuring certain critical alerts email directly to our HPE colleagues' pagers.

Checkmk is our first port of call for investigating issues reported by users, colleagues or stakeholders. The availability of an at-a-glance view of system status has been of great utility.

Checkmk is easily deployed to client servers - a Checkmk agent is deployed to the relevant node which conducts most checks when polled by the server. It can also run more heavyweight checks in the background. Polling is available via xinetd or systemd socket however this can also be configured to operate via ssh or any custom command. Deployment is available via rpm/deb and we have historically deployed the Checkmk agent to all management and login servers but not to compute nodes.

"Out of the box" Checkmk provides checks including:
\begin{itemize}
\item CPU, Memory, disk utilisation and load;
\item IPMI checks (fans, temperatures, voltages);
\item network interface status and statistics;
\item file system mounts;
\item individual processes can be assigned for monitoring (e.g. PBS mom or license servers); and
\item number of users logged in
\end{itemize}

It is also possible for the server to directly query clients via protocols such as SNMP. We have used this to quickly deploy monitoring for systems such as switches and tape libraries.

In addition to checks available by default, a variety of checks are available online to import. A number of checks have been imported over time including checks for monitoring some more specialised systems, such as the RAID controllers for a particular storage system.

Motivated in a large part by the data gathered by Checkmk, EPCC has deployed a Graphite\cite{b8} metrics server and a Grafana\cite{b9} analytics and visualisation server. This allows for metric based monitoring data gathered by Checkmk, as well as from other sources, to be combined and viewed in graphs and dashboards. This also supports greater visibility of data both within and beyond EPCC as a variety of stakeholders can be given access to custom dashboards to allow monitoring of data relevant to individual interests.

\subsection{Specialised checks}
One particular utility of Checkmk is the ease with which new monitoring items, or "checks", can be crafted and deployed. New checks can be deployed either as full blown plug-ins or as simple scripts, outputting health and metric data in the appropriate format when called. When scripts are deployed locally in this second method they will be automatically run by Checkmk once per minute.

Over time we have deployed a number of specialised checks in support of HPC services. These have included checks intended to monitor specialised HPC technologies as well as checks to detect recurring problems. Specialised checks deployed at EPCC include:
\begin{itemize}
    \item a check to monitor the health status of DDN controller servers;
    \item a check to monitor the status of GPFS clusters;
    \item a check to capture events observed by SELinux and AppArmor;
    \item a check to capture metric data regarding lustre server statistics;
    \item a check to monitor for the occurrence of unplaceable and orphan jobs in PBS Pro before these could impact system scheduling;
    \item a check to use cluster manager commands to capture compute node status on HPCM based systems; and
    \item a check to capture the health status of an Omnipath network using the Omni-Path Fabric Toolset.
\end{itemize}
These various checks were of significant benefit in the deployment and management of the systems for which they were implemented. As such, going into the ARCHER2 deployment, use of Checkmk, Graphite, Grafana and defining specialised checks were all key parts of our planning.

\section{Monitoring implemented during the ARCHER2 deployment}
\subsection{Overview of deployed infrastructure for ARCHER2}

\begin{figure}
  \centering
  \includegraphics[width=\linewidth]{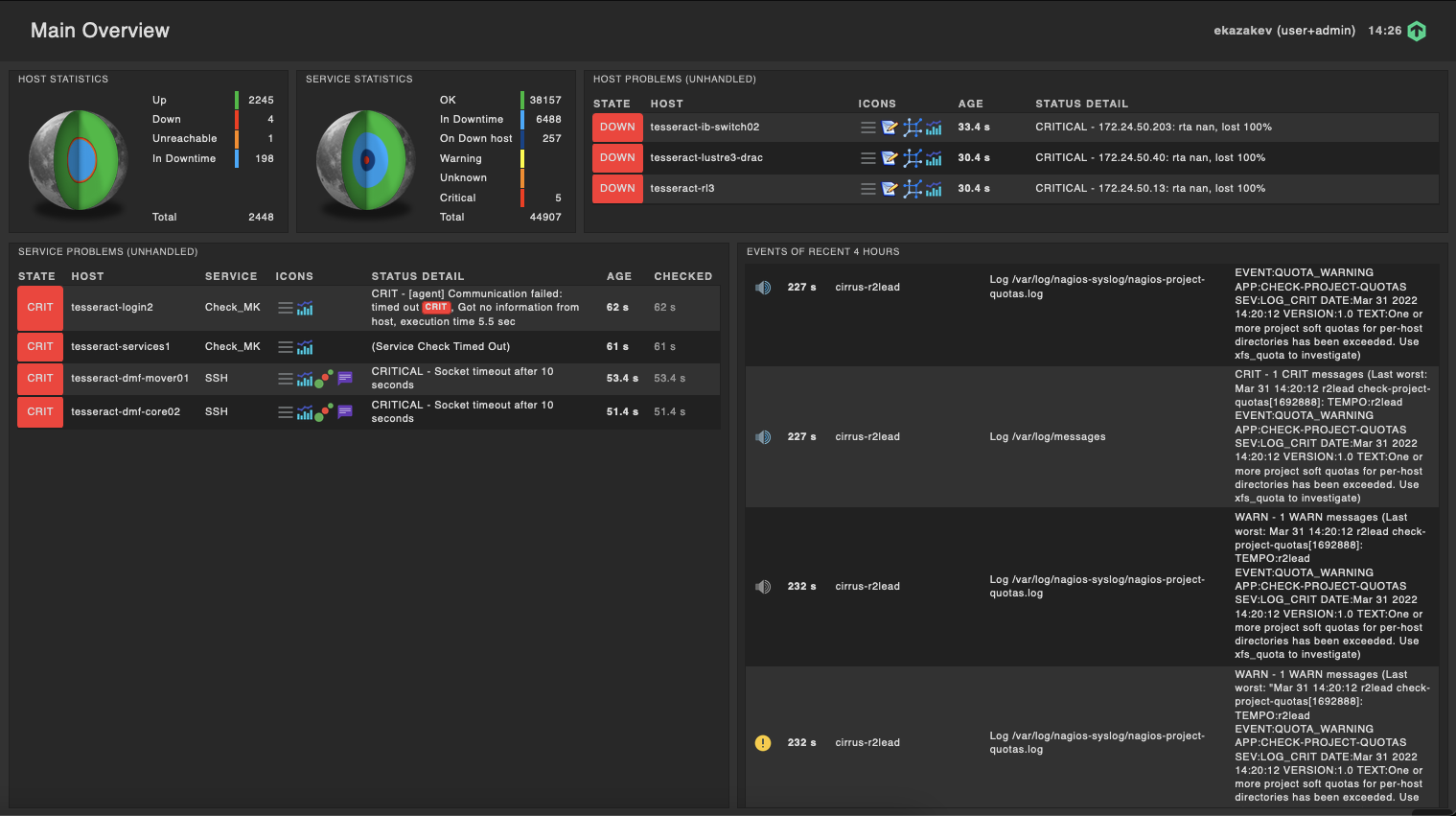}
  \caption{The front page for the EPCC central Checkmk server}
  \label{Panopticon_front_page}
\end{figure}

\begin{figure}
  \centering
  \includegraphics[width=\linewidth]{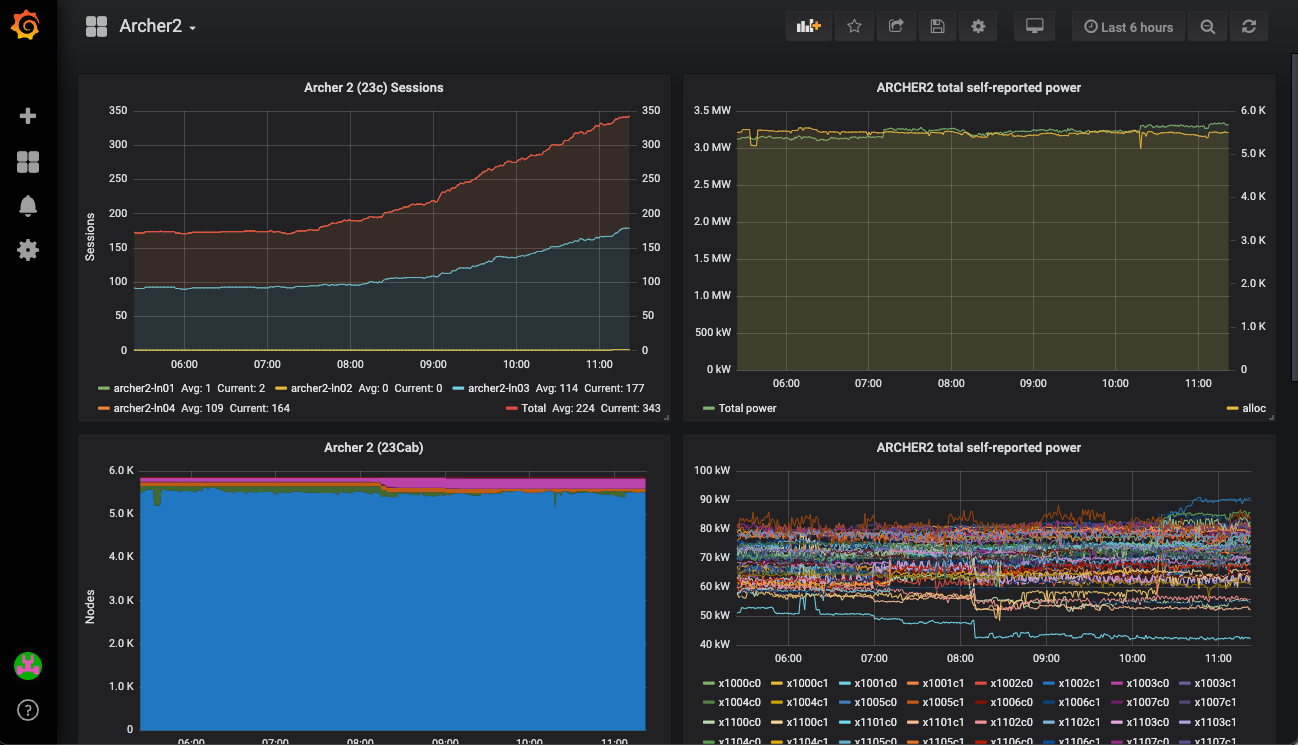}
  \caption{An overview page for a number of ARCHER2 graphs on Grafana}
  \label{Lens_ARCHER2_page}
\end{figure}

Each system or group of systems has a separate monitoring server that is controlled from the central Checkmk instance \cite{b1}. This approach provides many benefits, including:
\begin{itemize}
    \item There is almost no network communication between central and system specific hosts;
    \item One monitoring host failure does not affect the overall monitoring setup;
    \item It is easy to add and remove new monitoring servers.
\end{itemize} 
An outline of the monitoring setup for ARCHER2 is shown in Figure \ref{monitoring_setup_diagram}. Each monitored host has a Checkmk agent installed which communicates to the server via TCP. This agent collects various host health, performance metrics and posts these to the monitoring server. Once the monitoring data reach the Checkmk server it is further passed on to the Graphite graphing server which process data using "Carbon" daemons and stores it in  Graphite's specialised database \cite{b2}. In addition to the motivations listed previously, the default Checkmk metric storage engine struggles to perform appropriately when asked to display large number of metrics \cite{b3}. As discussed previously EPCC has three methods to access system status information. Firstly, all critical notifications are directly dispatched to appropriate personnel email inboxes; for example if login node DNS resolution fails, all system support staff get an email notification. There are then two graphical user interfaces accessible via web browser: a centralised Checkmk control centre that presents overview of all hosts, services, and checks (Figure \ref{Panopticon_front_page}); as well as a Grafana analytics and visualisation web application that pulls various metrics from the Graphite metrics server and presents them in the form of customisable and versatile graphs (Figure \ref{Lens_ARCHER2_page}).

As well as collecting the default set of data available from the Checkmk agent and redeploying some checks used elsewhere, we have deployed a number of new custom checks during the ARCHER2 rollout in response to emerging needs. These are discussed in more detail below. In each case, the check is deployed as a bash script placed into the appropriate directory (\url{/usr/lib/check_mk_agent/local}). Checks can however be deployed using any language supported by the host - the only requirement is that the output of the check be in the correct format \cite{b4}. Once a check has been deployed to the appropriate directory on a client, the Checkmk web interface on the server can be used to discover the new service.
\subsection{Power monitoring}
\begin{figure}
  \centering
  \includegraphics[width=\linewidth]{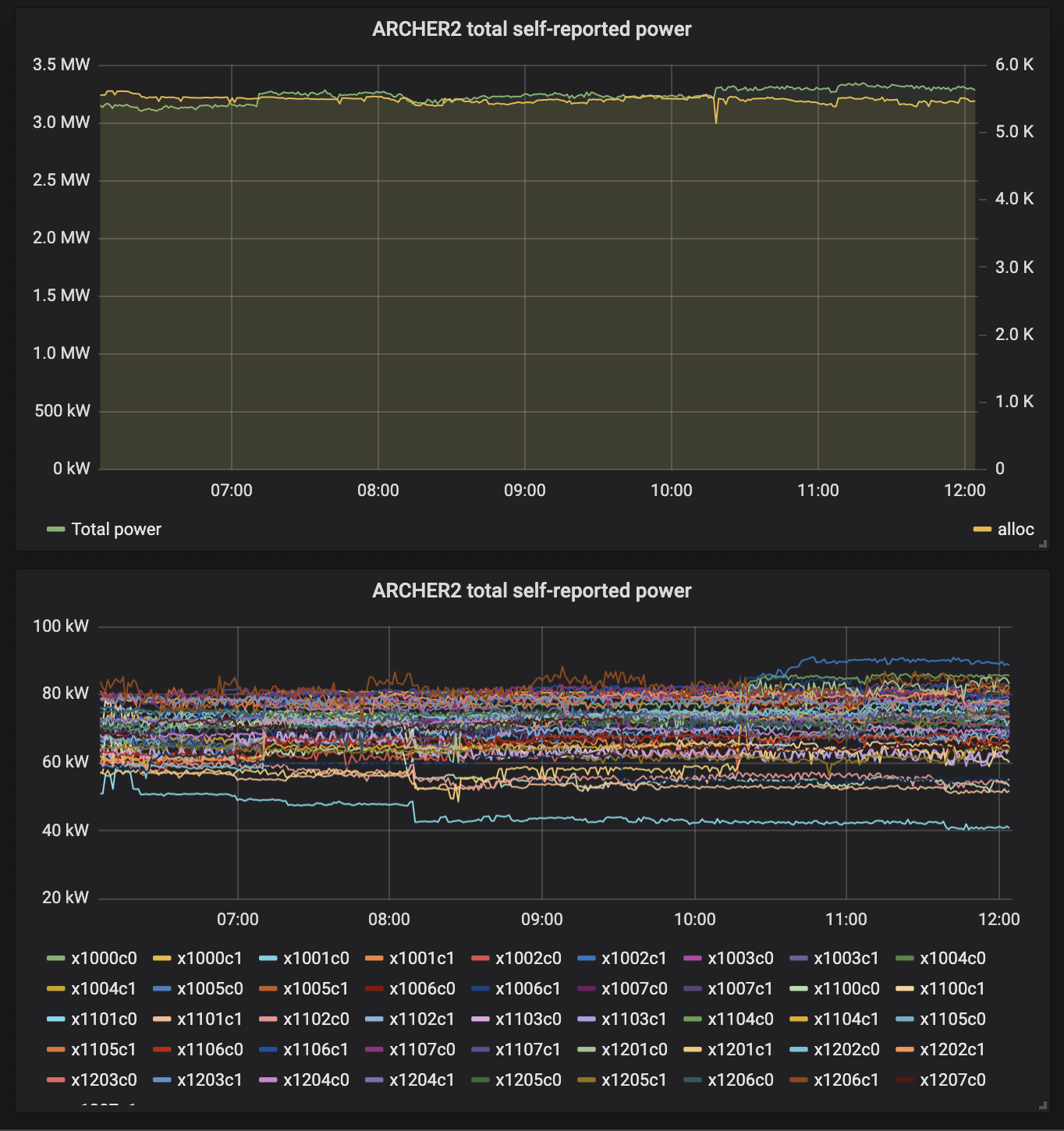}
  \caption{Graphs showing the whole system and per-cabinet power data gathered}
  \label{Lens_power}
\end{figure}
The ARCHER2 system is noticeably larger than it's predecessor and has a larger power profile. This profile sits at the maximum of what the Computer Room in which it sits was designed to support. As such there was a need to work carefully when the system was first brought into full testing and a requirement for a strong awareness of the power draw of the system at any time.

HPE identified that power data at a rectifier level was available via the cabinet controllers and automated the collection of this data at a relatively high granularity (every five seconds) to a local file on an admin node of the system. In order to pull this data into our monitoring setup, a new Checkmk local check was deployed.

This check:
\begin{itemize}
    \item uses pdsh to access each cabinet controller in turn;
    \item on each cabinet controller gathers power data found in \url{/var/volatile/cec/rectifiers} and stores this for analysis;
    \item iterates over the data to analyse power and voltage
    \item outputs the power draw on a per-cabinet basis;
    \item outputs the power draw on a whole system basis; and
    \item outputs the voltage on a per-rectifier basis.
\end{itemize}

There was no direct requirement to gather the voltage data however, given that there was no difficulty including it here, this was included against potential future need. Figure \ref{Lens_power} shows the graphing of the data gathered by this check.

\subsection{Node state monitoring}
\begin{figure}
  \centering
  \includegraphics[width=\linewidth]{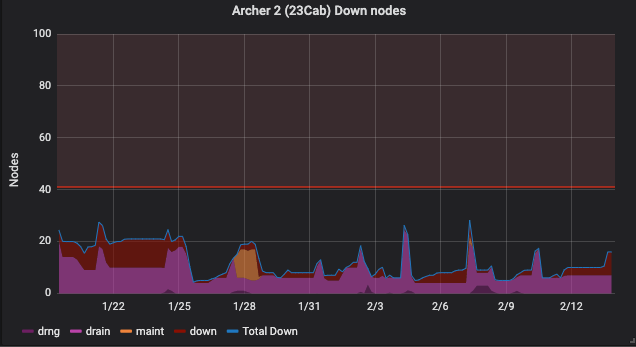}
  \caption{Graph showing the number of nodes in the various states considered "down" }
  \label{Lens_down_nodes}
\end{figure}
As deployment of the 23 cabinet system was taking place and a variety of issues were being troubleshooted, a requirement emerged for tracking the status of all compute nodes to provide an awareness of the state of the system at any given time. In order to gather this data, a check was scripted and deployed to the login nodes which assesses the state of the compute nodes via the Slurm scheduler.

This check does the following on each of the four login nodes:
\begin{itemize}
    \item runs "sinfo" and stores the output;
    \item pulls the names of the various partitions from the sinfo output;
    \item for each partition stores the number of nodes in each of the possible Slurm node states; and
    \item outputs the total counts for each node type on a per-partition basis.
\end{itemize}

One advantage of the approach taken here, with all partitions assessed and reported automatically, is that we have been able to implement this script against other systems using Slurm on-site without modification. A graph showing data gathered from this check can be seen in Figure \ref{Lens_down_nodes}. In this graph the total number of nodes in any of the states considered "down" is listed - the line shown represents the targeted node availability threshold.

This check is reported by each of the four login nodes. In order to have a single metric which is persistent regardless of which login nodes are up or down (so long as at least one is up) a "cluster host" was created within Checkmk which takes in the reporting from each login node and reports a single metric \cite{b5}.

\subsection{Login availability monitoring}
As part of EPCC's responsibilities there is a requirement to monitor the availability of the service. In order to support this monitoring a requirement was determined for a check to monitor the availability of the ARCHER2 login service. On the ARCHER2 service there is a single DNS record which serves all login nodes on a round-robin basis. In order to test this, the following setup was put in place:

\begin{itemize}
    \item a functional test user account was created;
    \item login access for this user was limited within the sshd config to the IP address of the Checkmk monitoring server;
    \item the test user account was added to the allow list in the sshd config for single factor (key based) access;
    \item credentials were put in place to allow ssh from the Checkmk user on the monitoring server to the login nodes; and
    \item a simple check script was deployed to the monitoring server which attempts to ssh to the login DNS address with the command "exit" and outputs the status of the login server based on the exit status of the attempted ssh.
\end{itemize}

\section{Impact of monitoring during the ARCHER2 deployment}
\subsection{Support for early service deployment and initial testing of the 23 cabinet system}
As with previous deployments of HPC services at EPCC, the system deployment team found early implementation and integration of monitoring extremely useful. This was in line with the experiences from other services described previously but a number of incidents are worth noting:
\begin{itemize}
    \item A number of problems were experienced with the provision of both internal and external DNS during the deployment of ARCHER2. Deploying a DNS resolution check to the login node allowed us to be rapidly alerted when problems occurred. This allowed prompt investigation and restoration of DNS.
    \item An incident occurred relating to the network providing a file system hosted elsewhere at the ACF data centre - the only original symptom of this was an inability for users to log into the system. Using Checkmk we were rapidly able to identify the origin of this issue as relating to the relevant file system and hence to the relevant network.
    \item In the early days of both the 4 cabinet and 23 cabinet systems a number of problems were experienced with the Slingshot High Speed Networks (HSN). The first indicator of this issue was often when Checkmk on the login node indicated a drop in the number listed for available Lustre LFS servers.
    \item We were able to quickly become aware of and begin troubleshooting of a memory leak on the login servers. Further as we were graphing all the data gathered we were able to assess the speed with which the leak was progressing and could reboot the login nodes at appropriate intervals until the problem was resolved.
\end{itemize}

\subsection {Support for system testing and benchmarking}

\subsubsection{Initial testing}
As discussed previously, ARCHER2 has a noticeably larger power profile than it's predecessor, ARCHER. ARCHER2 sits at the maximum of the design intent for the Computer Room in which it is hosted. As such, additional care was taken during the initial testing of ARCHER2 system.

In initial testing, using non-optimised High Performance Linpack (HPL) at 4,000-5,000 nodes, power use was initially monitored directly from figures gathered at the wall-level Power Distribution Units (PDUs) and via the Building Management System (BMS). The data gathered from these sources was found to be difficult to access, not as accurate as was preferred and not available to be accessed in a suitable graphed form. 

HPE identified that appropriate data was available via the cabinet controllers on the system and made this data available - as is described in the section above, this was integrated into our Checkmk monitoring and made available in graph form via Graphite and Grafana.

This provision, combined with verifying figures against those gathered from PDUs and the BMS, allowed us to build confidence that the system was operating correctly and safely at scale. We were able to profile the power draw of the system when operating at scale with benchmarks including HPL and the ARCHER2 procurement application benchmarks:  OpenSBLI, HadGEM3 (UK Met Office Unified Model), GROMACS and CASTEP.

Additionally, given that this data was available remotely, we were able to agree with HPE for their US teams to operate on the system at scale out-of-hours earlier in the life of the service than would otherwise have been possible. HPE's US team had access to the monitoring data and thresholds were agreed for power draw at which work would need to be stopped.

\subsubsection{HPL Benchmarking}
\begin{figure}[t]
  \centering
  \includegraphics[width=\linewidth]{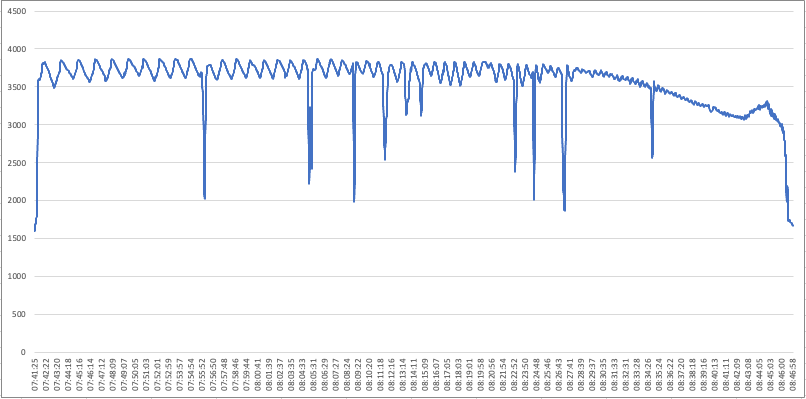}
  \caption{Graph showing the power draw (in kW) during an HPL run impacted by the power cycling issue. This ran on 5,500 nodes and achieved 16.8PF.}
  \label{HPL_Bad}
\end{figure}
\begin{figure}[t]
  \centering
  \includegraphics[width=\linewidth]{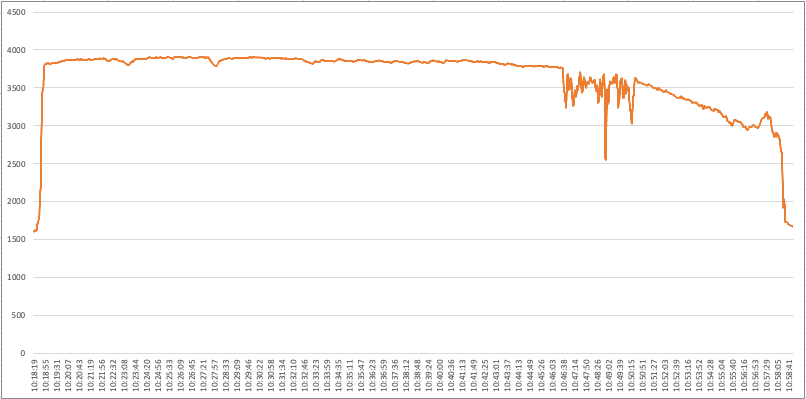}
  \caption{Graph showing the power draw (in kW) during an HPL run which was less impacted by the power cycling issue. This ran on 5,576 nodes and achieved 18PF}
  \label{HPL_Good}
\end{figure}
\begin{figure}[t]
  \centering
  \includegraphics[width=\linewidth]{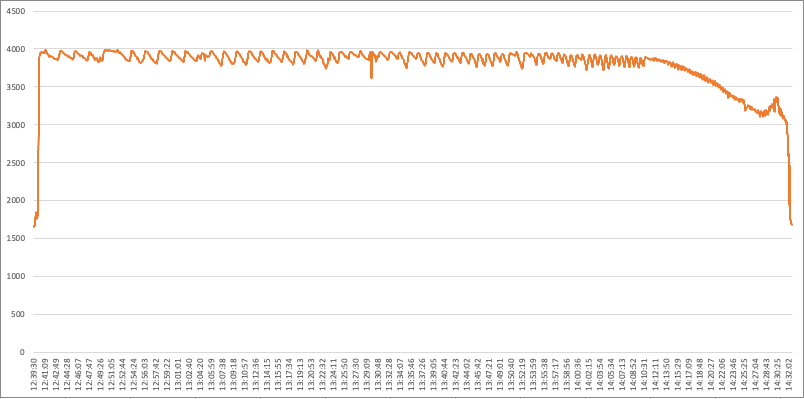}
  \caption{Graph showing the power draw (in kW) during the HPL run submitted to the Top 500. This ran on 5,600 nodes and achieved 19.5PF}
  \label{HPL_Final}
\end{figure}
The monitoring of power draw on the service was again useful during efforts to prepare an HPL benchmark suitable for submission to the Top 500 list. Over the course of a week, a number of attempts were made to produce a suitable result. A good number of runs were interrupted by node failures or problems with the HSN, however we were able to complete a number of runs. Note that the graphs shown in Figures \ref{HPL_Bad}-\ref{HPL_Final} have been generated from data gathered directly on the system. The granularity of data retained on Graphite is reduced over time to save disk space and full granularity for this data is no longer available at the time of writing.

It quickly became apparent that we were seeing "power cycling" behaviour on the system during these HPL runs, where power usage dropped suddenly for a short period of time. An example of a run impacted by this problem can be seen in Figure \ref{HPL_Bad}. In order to analyse this issue, single node HPL was run across the system and it was identified that certain nodes were performing persistently poorly. With these nodes drained, further testing showed that the problem was removed or reduced. An example of a run where this problem has been significantly reduced can be seen in Figure \ref{HPL_Good}.

Following this work, a number of runs were made to achieve the best available Figure for our Top 500 submission. Using the power monitoring we were able to quickly identify jobs impacted by the power cycling issue and then scan for and drain problem nodes. At the end of the week were able to achieve a score of 19.5PF which, when submitted, placed ARCHER2 at number 22 on the Top 500. The power profile of the submitted run is shown in Figure \ref{HPL_Final}.

\subsection{Automated contractual monitoring and system status website}
\begin{figure}[t]
  \centering
  \includegraphics[width=\linewidth]{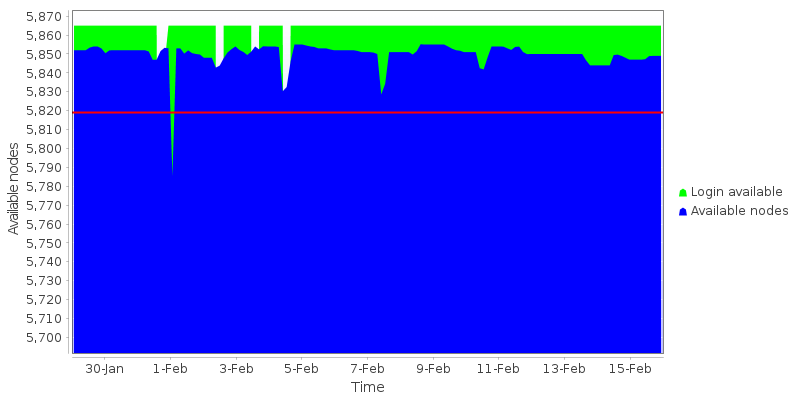}
  \caption{Graph showing the two elements of contractually define system availability. The number of available nodes is shown in blue, with the contractual threshold in red. The availability of the login service is shown in green.}
  \label{Contractual_Availability}
\end{figure}
We have also been able to make use of the monitoring data gathered beyond stakeholders in EPCC and HPE. In order to support UKRI (the funders) in their monitoring of the service over the acceptance trial period, a requirement emerged to prepare a single view which would encompass all attributes of the service relevant to the contractual monitoring of the service. This includes node availability, login availability and job failures. 

EPCC develops and operates a service management web application known as SAFE. Details of job failures, and other data relevant to job accounting, are uploaded to SAFE. SAFE is also used by users to create and manage their ARCHER2 accounts, and by project managers to allocate and report on project resources. Data on Graphite was exposed to the SAFE via a web API over HTTP. This allowed the SAFE to pull in relevant data relating to node availability and login availability gathered by the checks described previously.

Using this data any authorised stakeholder is able to generate a report in SAFE covering the contractual monitoring of the service for any given period. Critically, SAFE allows for fine-grained access control so only specifically authorised people can run these reports in SAFE. An example of the graph showing the login availability and compute node availability generated by this report is shown in Figure \ref{Contractual_Availability}.

In addition to stakeholders in EPCC, HPE and UKRI, this data is made available to the user community as a whole. Graphs of node availability are generated by the SAFE based on the data from Checkmk/Graphite and are presented on the ARCHER2 System Status web page\cite{b6}.
\section{Conclusions}
We have found that live monitoring and graphing makes an extremely valuable contribution to service management. Furthermore this value often presents itself in unexpected ways - we would not have anticipated when first deploying Checkmk and Graphite/Grafana that we would see benefits such as those during the testing and benchmarking of ARCHER2 or that we would be able to implement contractual monitoring using these technologies.

The ability to rapidly and flexibly deploy new checks in response to emerging events and requirements is also of particular value - and an imperfect check implemented rapidly is often superior to an ideal check which might require greater deployment time.

It is also clear that automating the contractual monitoring of a service can be extremely valuable. This helps us to assure service partners, funders and users that system is working. This has been particularly important given the delayed start to ARCHER2.

We finally note that ARCHER2 has now been in full service for some months with in excess of 2,500 active users and utilisation on the order of 90\%. We consider automated monitoring to have been key in making this possible.

\section{Future work}
We are interested in further developing our automated monitoring capabilities going forward. Potential improvements include integrating:

\begin{itemize}
    \item log analysis;
    \item Slingshot error feeds;
    \item per-job lustre statistics and
    \item data driven intrusion detection.
\end{itemize}

We are also interested in making the data we collect more generally available to our user community.

We would be pleased to coordinate with other sites who use or are interested in using Checkmk for HPC service monitoring and are happy to share our experience.


\begin{thebibliography}{00}
\bibitem{b7} \url{https://checkmk.com/}
\bibitem{b8} \url{https://graphiteapp.org/}
\bibitem{b9} \url{https://grafana.com/}
\bibitem{b1} \url{https://docs.Checkmk.com/latest/en/distributed_monitoring.html}
\bibitem{b2} \url{https://graphite.readthedocs.io/en/latest/carbon-daemons.html}
\bibitem{b3} \url{https://wiki.lustre.org/Check_MK/Graphite/Graphios_Setup_Guide}
\bibitem{b4} \url{https://docs.Checkmk.com/latest/en/localchecks.html}
\bibitem{b5} \url{https://docs.Checkmk.com/latest/en/clustered_services.html}
\bibitem{b6} \url{https://www.archer2.ac.uk/support-access/status.html}
\end{thebibliography}
\end{document}